\documentclass[journal=jpc,manuscript=article]{achemso}

\usepackage[version=3]{mhchem} 
\usepackage{graphicx}
\usepackage{amsmath, mathtools}
\usepackage{array}
\usepackage{xcolor}

\author{Jan-Niklas Boyn}

\affiliation[The University of Chicago]{Department of Chemistry and The James Franck Institute, The University of Chicago, Chicago, Illinois 60637 USA}

\email{jboyn@uchicago.edu}
\author{Lauren E. McNamara}

\affiliation[The University of Chicago]{Department of Chemistry, The University of Chicago, Chicago, Illinois 60637 USA}

\author{John S. Anderson}

\affiliation[The University of Chicago]{Department of Chemistry, The University of Chicago, Chicago, Illinois 60637 USA}

\author{David A. Mazziotti}

\affiliation[The University of Chicago]{Department of Chemistry and The James Franck Institute, The University of Chicago, Chicago, Illinois 60637 USA}

\email{damazz@uchicago.edu}
\affiliation[The University of Chicago]{Department of Chemistry and The James Franck Institute, The University of Chicago, Chicago, Illinois 60637 USA}

\title[]{Interplay of Electronic and Geometric Structure Tunes Organic Biradical Character in Bimetallic Tetrathiafulvalene Tetrathiolate Complexes}

\DeclareUnicodeCharacter{2212}{-}
\DeclareUnicodeCharacter{2005}{ }

\begin{document}

\begin{tocentry}
    \centering
    \includegraphics{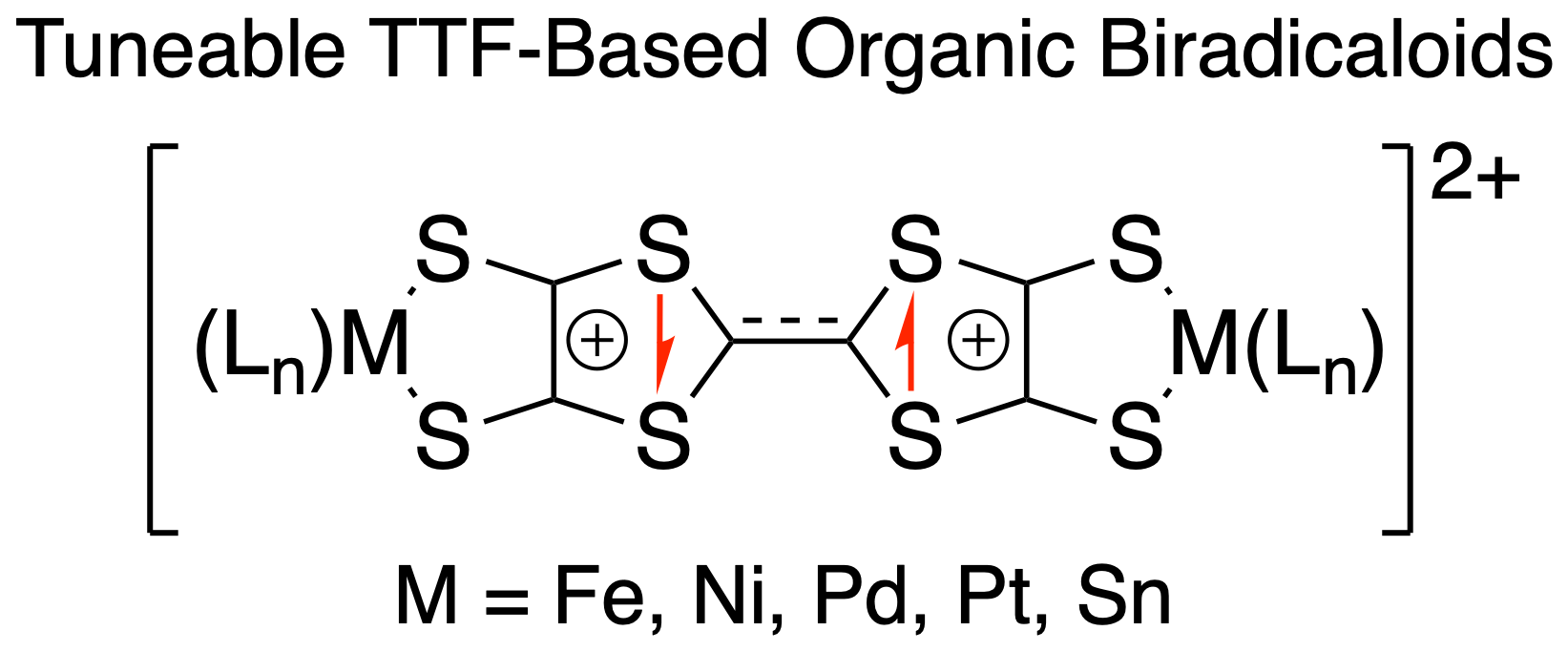}
\end{tocentry}


\begin{abstract}
The synthesis and design of organic biradicals with tunable singlet-triplet gaps has become the subject of significant research interest, owing to their possible photochemical applications and use in the development of molecular switches and conductors. Recently, tetrathiafulvalene tetrathiolate (TTFtt) has been demonstrated to exhibit such organic biradical character in doubly ionized bimetallic complexes. In this article we use high-level {\em ab initio} calculations to interrogate the electronic structure of a series of TTFtt-bridged metal complexes, resolving the factors governing their biradical character and singlet-triplet gaps. We show that the degree of biradical character correlates with a readily measured experimental predictor, the central TTFtt C-C bond length, and that it may be described by a one-parameter model, providing valuable insight for the future rational design of TTFtt based biradical compounds and materials.
\end{abstract}

\section{Introduction}
Since first being reported in the early 1970s\cite{TTF1, TTF2, TTF3}, tetrathiafulvalene (TTF) has been established as a valuable building block in multiple areas of chemistry and materials science\cite{RevLigands, SwitchProcesses, REVOrgMet, RevMacro, REV5, REV6, REV-OrgElec, REVTTFC60,REVTTF-Donor-Aceptor,REVTTF-H-Halogen-Bonding-MCond}. This includes the use of TTF and its derivatives in organic conductors\cite{REVTTF-H-Halogen-Bonding-MCond, REV-OrgElec, REVOrgMet, NiTTFtt}, metal or covalent organic frameworks\cite{TTFMOF-ChargeMobility, 3DCOF, MOFpi}, molecular switches\cite{SwitchProcesses, FeSwitch}, coordinating ligands in transition metal and lanthanide chemistry\cite{RevLigands, DyYbLigand, RevLigandLn, LnTTF_LumSMM, LigandRevBuildingBlocks, SMMLigands, TTFPLigands} and macrocyclic and supramolecular structures\cite{RevMacro, Containers, Shuttles}. The versatility in its applications arises from its tunable electronic structure, namely its extended $\pi$-system that yields low-lying excitations and accessible $1+$ and, more recently, $2+$ oxidation states\cite{Transmetalation, NiTTFtt}. Promising applications of TTF-derived compounds include their roles in the development of single molecule magnets (SMM)\cite{SMMLigands, LnTTF_LumSMM}, molecular transistors\cite{TTFTransistors}, as well as materials for light harvesting and solar energy conversion\cite{REVTTFC60}. \\

Recently, tetrathiafulvalene tetrathiolate (TTFtt)\cite{TTFtt, TTFttLigand} has found application as a bridging ligand in bimetallic transition metal complexes\cite{TTFtt_Bimetallics1, TTFtt_Bimetallics2, Transmetalation}. Here TTFtt may be utilized as a formal TTFtt$^{4-}$, TTFtt$^{3-}$ or TTFtt$^{2-}$ ligand, where the TTF-core is in its neutral, $1+$ or $2+$ state, respectively, \textcolor{black}{accessed via oxidation of the neutrally charged bimetallic parent compounds\cite{Transmetalation}}. While radical TTF$^{1+}$ ligands are well studied and have found applications in the development of SMM\cite{SMMLigands, LnTTF_LumSMM}, the TTFtt$^{2-}$ ligand has only recently been shown to yield significant TTF-based organic biradical character as a bridging ligand in transition metal bimetallic complexes\cite{FeSwitch}. Further TTFtt$^{2-}$ bridged complexes containing different metal centers have been synthesized and demonstrated to display high photoluminescence quantum yield (PLQY)\cite{TMTTFtt-Lumi}, as well as high conductivity in amorphous, glassy coordination polymers\cite{NiTTFtt}. \\

In this article we perform high-level {\em ab initio} electronic structure calculations on a series of TTFtt$^{2-}$-bridged bimetallic complexes. Since singlet biradical states are an example of a problem dictated by strong correlation---arising in systems containing degenerate or near-degenerate orbitals that cannot be described by a single Slater determinant, they require computational treatment with a multi-reference theory. Here, we use the complete active space self consistent field (CASSCF) method, allowing us to capture the singlet biradical character of these complexes accurately. Additional accuracy in the calculation of their singlet-triplet gaps is achieved via the inclusion of dynamic correlation effects with $N$-electron valence state perturbation theory (NEVPT2)\cite{NEVPT2} and the anti-Hermitian contracted Schr\"odinger equation (ACSE)\cite{ACSE, ACSE2, SA-ACSE, MRACSE, ESACSE, OSACSE, PhotoExACSE, ACSE,CQE,QACSE,QCACSE,CoDimer} theory with cumulant reduced-density-matrix reconstruction\cite{schwinger, CSE}. Our calculations allow us to elucidate the interplay between both structural and electronic effects governing the trends in the biradical character and T-S gaps of the bimetallic TTFtt$^{2-}$ compounds. As tuneable biradical character\cite{diradicals-rev} is a particularly desirable property for the design of molecular switches\cite{FeSwitch, NonCovParaSwitch, photoresponsive}, qubits\cite{qubit,qubits,qbitdesign} and spintronics,\cite{spintronics,organicspintronics,structure-spinspin,biradical-acenes} our work will provide valuable insight for future rational design of TTFtt-based molecules and materials. \\

\begin{figure*}[h]
    \centering
    \includegraphics[scale=0.25]{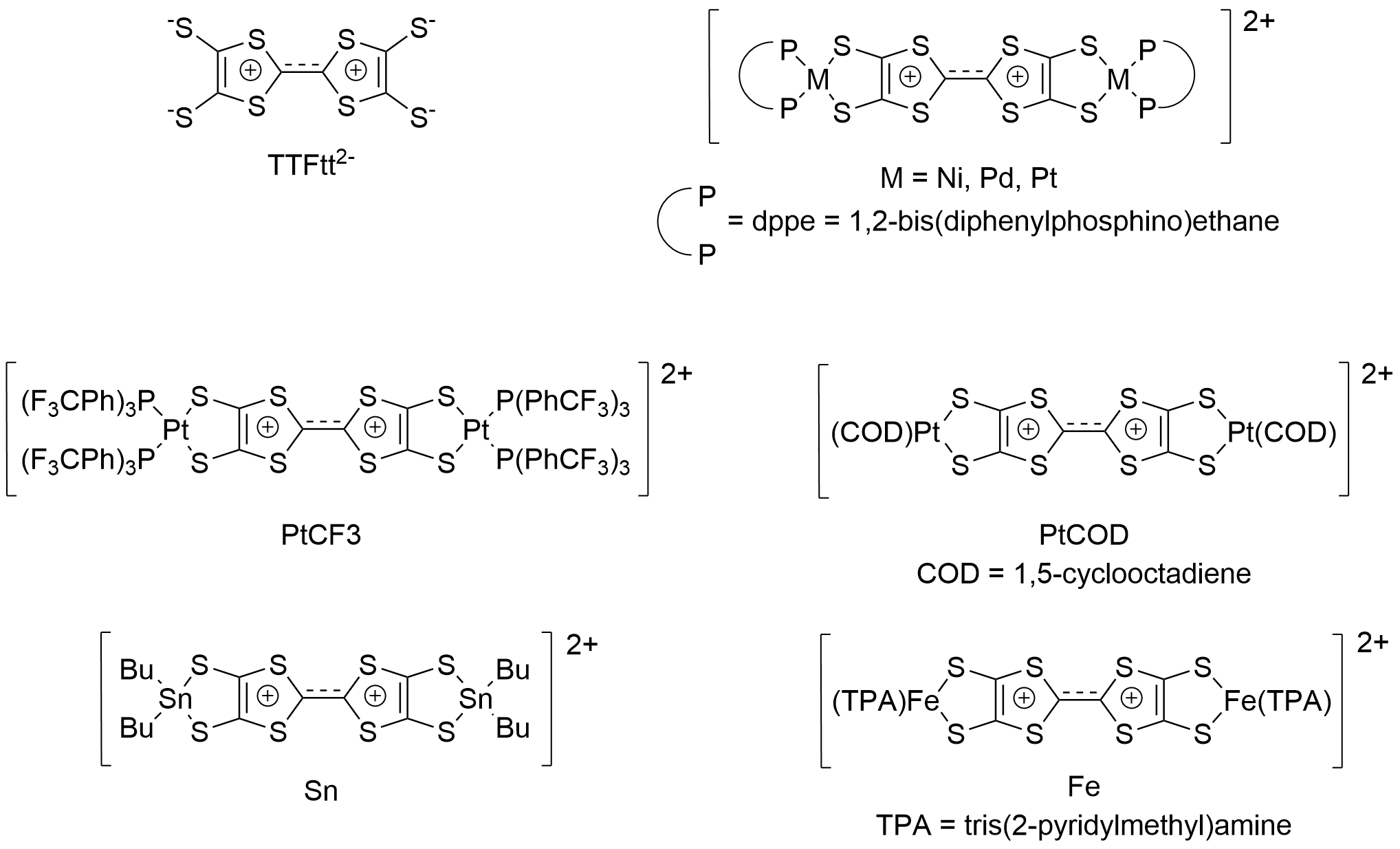}
    \caption{Overview of the molecules considered. We refer to the various complexes in the text by the identity of the relevant metal center, and in the case of the two additional platinum complexes with the additional identity of the terminal ligands, i.e. PtCOD and PtCF3.}
    \label{fig:compounds}
\end{figure*}

\section{Methods}
The accurate resolution of the singlet biradical character present in the studied systems requires the use of multi-reference methods.  Here, we performed [4,4] active space CASSCF calculations with NEVPT2 as implemented in PySCF\cite{pyscf, PySCFCASSCF, Pyscf-nevpt2}, while larger CASSCF calculations with active spaces as large as [24,24] were performed with the variational 2-RDM (V2RDM) method\cite{SDP,SDP2,V2RDM,V2RDM2,V2RDMT2,V2DualCone,VOxo} as implemented in the Maple Quantum Chemistry Package\cite{Maple, QCP}. Additional post-CAS calculations were undertaken on the TTFtt ligand using the ACSE method with the Maple Quantum Chemistry Package, which when seeded with an initial guess from a multi-reference CAS calculation, allows the resolution of the on-top dynamic correlation with an accuracy comparable to that of coupled cluster with singles, doubles and perturbative triples (CCSD(T))\cite{ACSE, ACSE2, SA-ACSE}. Geometry optimizations and frequency calculations were performed using density functional theory (DFT) with the B3LYP\cite{B3LYP} and MN15\cite{MN15} functionals as implemented in Gaussian 16 Revision A.03\cite{g16}. The Pople basis sets 6-31G and 6-31G*\cite{basis,basis2} were used to treat the light atoms H, C, N, P, S, Fe, Ni while the heavy atoms Sn, Pd, Pt were treated using the LANL2DZ\cite{LANL2DZ} and LANL2TZ\cite{LANL2TZ} basis sets. DFT calculations for the fragment analysis were performed with the larger def2-TZVP basis set\cite{def2}. Orbital density plots were created using the VESTA program\cite{VESTA}. \\

\section{Discussion \& Results}

\begin{figure}[h]
    \centering
    \includegraphics[scale=0.33]{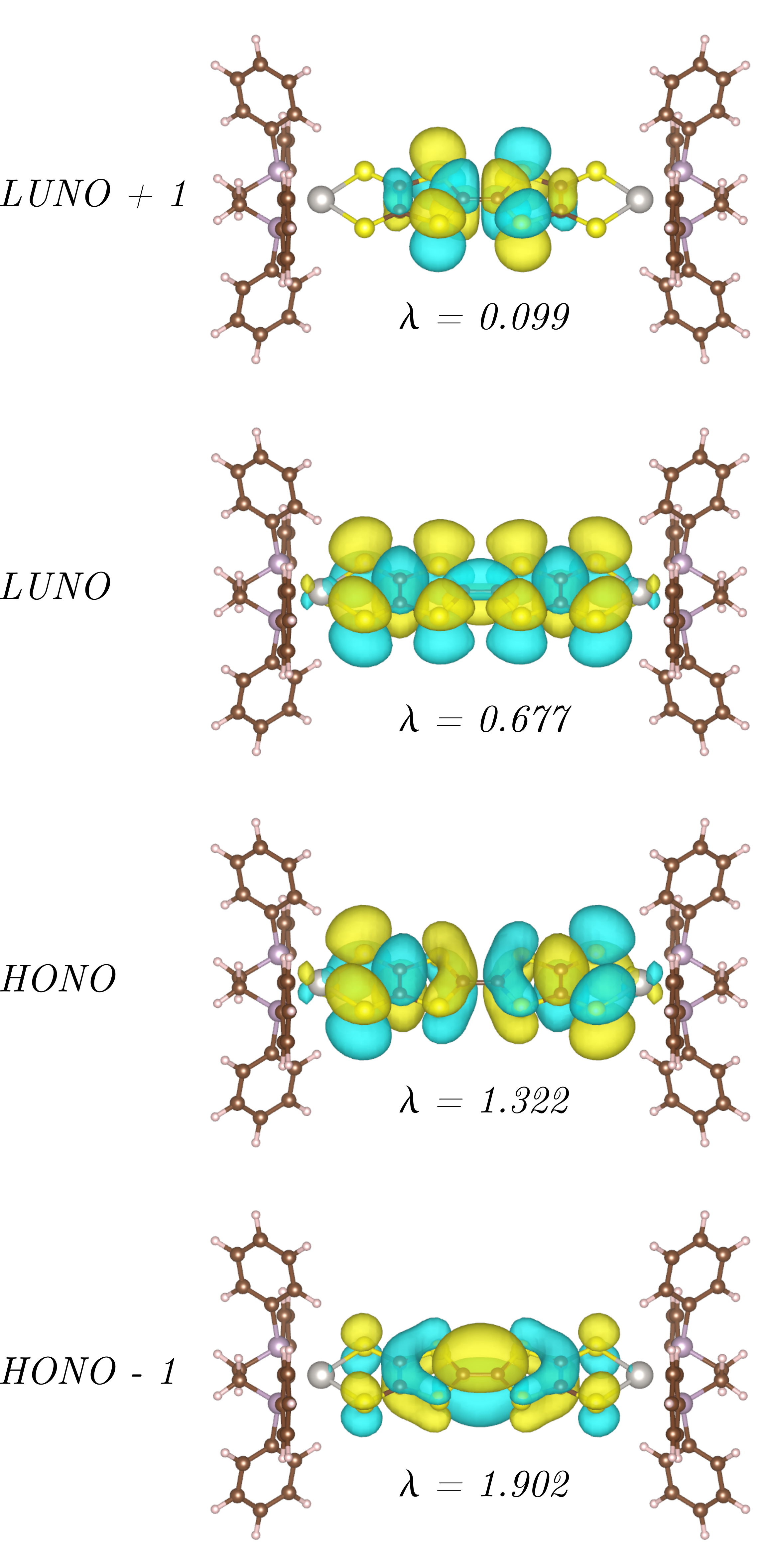}
    \caption{Frontier natural orbitals of the \textcolor{black}{[PtTTFtt]$^{2+}$} system obtained with [4,4] CASSCF and LAN2LDZ/6-31G basis sets.}
    \label{fig:V2NOs}
\end{figure}

To resolve any trends in the biradical character and triplet-singlet (T-S) gaps of bimetallic TTFtt$^{2-}$ bridged complexes, we investigate the electronic structure of a series of complexes which have been recently reported (Pt\cite{TMTTFtt-Lumi}, PtCF3\cite{TMTTFtt-Lumi}, Pd\cite{TMTTFtt-Lumi}, Ni\cite{Transmetalation}, Sn\cite{Transmetalation}, Fe\cite{FeSwitch}), as well as the newly proposed \textcolor{black}{and yet to be experimentally characterized} PtCOD. These compounds are of particular interest in the development of optically addressable molecular qubits or switches and a recent experimental and computational investigation by the authors has shown them to exhibit fluorescent behavior driven by a $\pi^* \rightarrow \pi$ transition of the TTFtt ligand\cite{TMTTFtt-Lumi}. The chemical structures of all investigated compounds are displayed in Figure \ref{fig:compounds}. \textcolor{black}{We note that while all complexes considered in this study are doubly charged cations they are referred to solely by the identity of the metal centers and ligands as indicated in Figure \ref{fig:compounds} with their charges being omitted in the following sections of this article.}\\

First, broken symmetry (BS) DFT geometry optimizations were performed with the B3LYP functional in combination with the LANL2DZ basis set for Pt, Pd, Sn and the 6-31G* basis set for all remaining atoms, relaxing the geometry from the one obtained in the solid state via single crystal X-ray diffraction (XRD)\cite{Transmetalation,FeSwitch,TMTTFtt-Lumi}. \textcolor{black}{The use of unrestricted Kohn-Sham DFT calculations, allowing them to break the spin-symmetry is essential for the accurate capture of the geometry and properties of systems displaying multi-reference character\cite{BSopt,BSopt2}}. The DFT optimizations were followed by high-level CASSCF and NEVPT2 calculations allowing us to capture accurately the effects arising from both the multi-reference and biradical character and dynamic correlation. A minimal active space comprising 4 electrons distributed in 4 spatial orbitals, [4,4], was chosen for the CASSCF calculations as the strong correlation in these systems has been shown to be limited to two frontier natural orbitals, largely comprised of TTFtt-based $\pi$ orbitals that give rise to the biradical character. The active-space natural orbitals of the \textcolor{black}{[PtTTFtt]$^{2+}$} system are \textcolor{black}{displayed} in Figure \ref{fig:V2NOs}, \textcolor{black}{showing the fractional occupation of the highest occupied natural orbital (HONO) and lowest unoccupied natural orbital (LUNO), as well as the lack of significant fractional occupation in the HONO - 1 and LUNO + 1}. This choice of a limited CASSCF active space is supported by \textcolor{black}{initial,} larger V2RDM CASSCF calculations performed on the PtTTFtt system in its XRD geometry using active spaces as large as [24,24], showing no significant changes to the predicted occupations of the HONO and LUNO, or the T-S gap (data shown in SI Table S1). The CASSCF and NEVPT2 calculations utilize the LANL2DZ basis set with its effective core potential for the heavy atoms Pt, Pd, and Sn and the 6-31G basis set for all remaining atoms. \textcolor{black}{While the use of the larger 6-31G$^*$ basis set, which results in a prohibitively large number of basis functions for all but the smallest compounds studied, yields a slight increase from 19.19 kcal/mol to 20.06 kcal/mol in the calculated T-S gap in PtCOD, changes to natural occupation numbers (NON) and $\Delta E_{T-S}$ are minor (see SI Table S2), meaning trends will likely be recovered correctly.}  \\

\begin{table*}[h]
    \centering
    \begin{tabular}{cccccccc}
        & Fe & Pt & PtCF3 & Ni & Pd & Sn & PtCOD \\
        \hline
        R(C-C) XRD & 1.366 & 1.399 & 1.411 & 1.411 & 1.421  & 1.437 & \\
        R(C-C) BS & 1.371 & 1.386 & 1.388 & 1.390 & 1.400 & 1.406 & 1.409 \\
        $S^2$ & 0.797 & 0.541 & 0.514 & 0.455 & 0 & 0.175 & 0 \\
        \textcolor{black}{$\Delta E_{\text{T-S,DFT}}$} & 2.28 & 3.99 & 5.47 & 6.17 & 5.61 & 9.35 & 6.86\\
        $\Delta E_{\text{T-S,CAS}}$ & 1.14 & 2.50 & 3.12 & 3.50 & 3.39  & 7.77 & 6.09 \\
        $\Delta E_{\text{T-S,NEVPT2}}$ & 2.84 & 6.67 &   & 9.59 & 8.88 & 19.19 & 16.55 \\
        $\lambda_{\text{LUNO, S}}$ & 0.796 & 0.677 & 0.630  & 0.610 & 0.615 & 0.359 & 0.451 \\
        $\lambda_{\text{HONO, S}}$ & 1.203 & 1.322 & 1.370  & 1.390 & 1.384 & 1.643 & 1.550 \\
    \end{tabular}
    \caption{Data for the DFT optimized structures. R(C-C) is the length of the central TTF C-C bond \textcolor{black}{in \AA ${}$} with R(C-C) BS obtained from broken-symmetry singlet DFT geometry and R(C-C) XRD obtained from experimental XRD data\cite{Transmetalation,TMTTFtt-Lumi,FeSwitch}, $S^2$ is the unrestricted singlet DFT spin contamination in the optimized geometry, $\Delta E_{\text{T-S}}$ denotes the triplet-singlet energy gap \textcolor{black}{in kcal/mol}, and $\lambda_{\text{HONO}}$ and $\lambda_{\text{LUNO}}$ are the occupation numbers of the HONO and LUNO, respectively.}
    \label{tab:DFTopt}
\end{table*}

Table \ref{tab:DFTopt} displays the vertical T-S gaps for the experimentally surveyed systems and the occupations of their frontier natural orbitals (NOs) based on the BS singlet DFT geometry. The CASSCF/NEVPT2 calculations reveal a significant amount of strong correlation and subsequent biradical character that varies across the different metal centers and ligands. The Fe complex shows the lowest triplet-singlet gap of 2.84 kcal/mol and the correspondingly greatest degree of biradical character with HONO and LUNO occupations of 1.20 and 0.80, respectively. The multi-reference character decreases along the series of \textcolor{black}{Fe $>$ Pt $>$ PtCF$_3$ $>$ Pd $>$ Ni $>$ PtCOD $>$ Sn}, with HONO and LUNO occupations declining to 1.64 and 0.36 in the case of Sn, while the T-S gaps follow the inverse trend and increase across the series to 19.19 kcal/mol in Sn. The trends observed in the NEVPT2 calculations are mirrored by the T-S gaps resolved with CASSCF alone; however, the inclusion of dynamic correlation effects leads to significant stabilization of the singlet state compared to the triplet and hence, the gaps predicted by NEVPT2 are of greater magnitude than those calculated from CASSCF.  \textcolor{black}{The values calculated with single-reference BS DFT and the B3LYP functional generally lie between those obtained with the multi-reference methods, recovering the same trends.}\\

Inspection of the bond distance between the central TTF C atoms, R(C-C), which follows the trend Fe $<$ Pt $<$ PtCF$_3$ $<$ Ni $<$ Pd $<$ Sn $<$ \textcolor{black}{PtCOD} in the optimized BS geometry, reveals a general correlation between said distance and the calculated $\Delta E_{\text{T-S}}$ and HONO and LUNO occupations. \textcolor{black}{The XRD data displays the identical ordering of R(C-C) in the surveyed compounds, however,} with the exception of Fe there is a general shift to longer R(C-C) lengths in the experimental solid state geometry which suggests that packing and intermolecular interactions arising from the observed $\pi$ stacking may lead to some minor structural and electronic changes. \textcolor{black}{Inspection of the BS DFT $S^2$ values furthermore reveals greater spin contamination as R(C-C) becomes smaller and the system becomes displays more multi-reference character, which is captured by the unrestricted DFT geometry optimizations.} \\

Analysis of the frontier NOs reveals that the unpaired electron density is heavily localized in two orbitals of the TTFtt based $\pi$-system, with only minimal involvement of metal-centered orbitals. A molecular-orbital diagram for the PtTTFtt$^{2+}$ complex showing the HONO-1 through LUNO+1 orbitals is displayed in Figure \ref{fig:V2NOs}. Additionally, inspection of orbital symmetries reveals the HONO to be of $\pi$-antibonding character across the central TTF C-C bond, while the LUNO is of bonding character. The length of the central TTF C-C bond shows variations that closely follow the observed trends in the CASSCF/NEVPT2 calculations: the Fe LT complex displays the shortest C-C distance in both XRD and DFT structures and simultaneously the largest occupation of the $\pi$-bonding LUNO, while the Sn analogue displays the longest C-C distance and the greatest $\pi$-antibonding HONO occupation.  \\

\subsection{Biradical Character Driven Structural Changes to TTFtt}
\begin{figure*}[h!]
    \centering
    \includegraphics[scale=0.7]{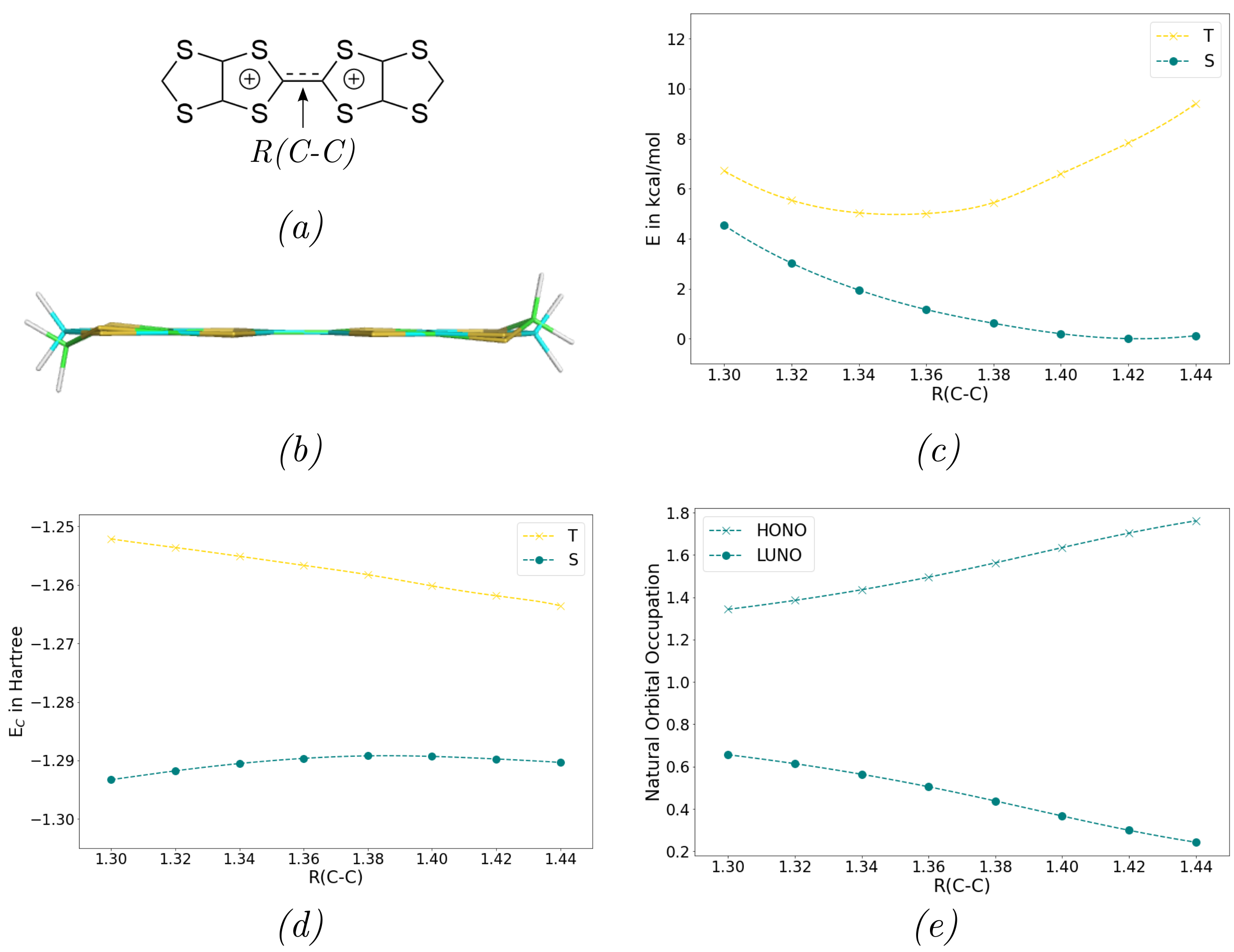}
    \caption{(a): Structure of the capped TTFtt$^{2+}$ molecule with the central C-C bond, which presents the scan coordinate indicated. (b): Overlay of the triplet and closed-shell singlet structures at their respective equilibrium geometries with the CH$_2$ group of the triplet geometry shown in blue and the CH$_2$ group of the singlet geometry in green. (c): A plot of the ACSE energy of the singlet and triplet state energies as a function of the central C-C bond length. Geometries at each point are obtained with constrained optimization with the respective multiplicity and the singlet uses broken symmetry. (d): ACSE correlation energies of the singlet and triplet as a function or R(C-C). (e): The HONO and LUNO occupations of the singlet as a function of R(C-C).}
    \label{fig:Scan}
\end{figure*}

To further investigate how structural changes correlate with the observed biradical character in TTFtt-bridged bimetallics, we performed calculations on CH$_2$ capped TTFtt$^{2+}$, scanning over the central TTF C-C bond (shown in Figure \ref{fig:Scan} (a)). The smaller size of the capped TTFtt$^{2+}$ allows us to perform high-level ACSE calculations to resolve accurately the effects of both dynamic and multi-reference correlation. As the strong correlation is limited to a set of two frontier $\pi$ orbitals, the ACSE was seeded with minimal active space [4,4] CASSCF calculations. We utilized the 6-31G basis set and applied the frozen-core approximation. \\

We investigate C-C distances ranging from 1.30 \AA \, to 1.48 \AA \, in steps of 0.02 \AA. For each frozen TTF C-C bond length, the geometry was optimized at the MN15/6-31G* level of theory and both the broken-symmetry singlet and triplet surfaces were evaluated. Figure \ref{fig:Scan} (c) displays the ACSE electronic energy as a function of the C-C bond distance. The data reveals that the structure of TTFtt$^{2+}$ is strongly dictated by its biradical and multi-reference character. The capped TTFtt$^{2+}$ structure shows an equilibrium bond distance of 1.42 \AA \, in the singlet state, which decreases to 1.34 \AA \, in the triplet state. Figure \ref{fig:Scan} (b) displays an overlay of the equilibrium singlet and triplet geometries, showing a clear out-of-plane bending of the terminal S-CH$_2$-S cap reducing its overlap with the TTF $\pi$ system as the biradical character is reduced and the system moves towards being closed-shell. A reduction in R(C-C) results in greater spin contamination in the singlet state and a reduction in the out-of-plane bending. In line with the respective equilibrium bond distances, the vertical T-S gap increases as the bond is stretched.\\

Further investigation of the total electronic correlation energy (defined as $E_c = E_{\text{ACSE}} - E_{\text{HF}}$, where $E_{\text{ACSE}}$ is the ACSE energy and $E_{\text{HF}}$ is the Hartree-Fock energy), and HONO and LUNO occupations, which are displayed in Figures \ref{fig:Scan} (d) and (e), respectively, shows a decrease in the magnitude of the correlation energy and a consequent loss of biradical character as the singlet becomes more closed-shell with increasing R(C-C), with the HONO occupation rising from 1.34 at 1.30 \AA \, to 1.76 at 1.48 \AA \, for the broken-symmetry geometry. This trend in the frontier NONs is mirrored by the DFT spin contamination which is reduced with increasing R(C-C) and reaches zero in the long C-C limit of 1.48 \AA.  \\

This data strongly suggests that the measurement of the TTF C-C distance by XRD can provide a valuable diagnostic tool in the determination of the degree of biradical character in TTF-based compounds. It may also provide a measure allowing for the comparison or estimation of T-S gaps which scale with increasing C-C distance. \\

\subsection{Fragment Orbital Analysis}
\begin{figure*}[h!]
    \centering
    \includegraphics[scale=0.30]{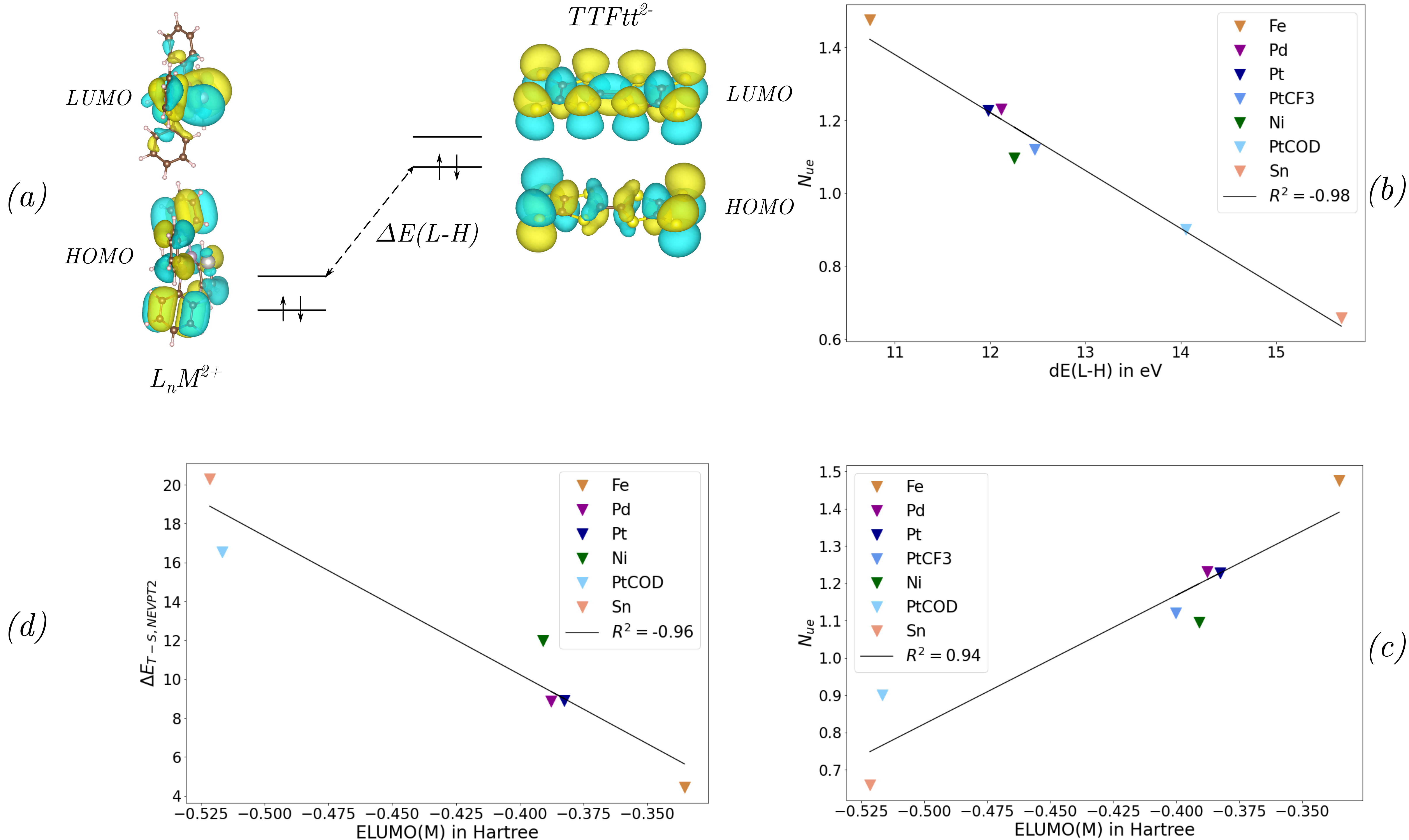}
    \caption{Plots showing the scaling of the biradical character with the fragment MO energies in the different TTFtt bridged systems. (a): Diagram of the frontier orbitals of the metal and TTFtt$^{2-}$ fragments, illustrating the interaction between the metal fragment LUMO and the TTFtt$^{2-}$ HOMO. (b): A plot of the biradical character, defined $N_{ue} = 2-\lambda_{HONO} + \lambda_{LUNO}$, against energy gap between the metal fragment LUMO and the TTFtt$^{2-}$ HOMO, $\Delta E_{L-H}$. (c): $N_{ue} = 2-\lambda_{HONO} + \lambda_{LUNO}$ plotted against the energy of the metal fragment LUMO. (d): The NEVPT2 T-S gap, $\Delta E_{T-S} \text{NEVPT2}$ of the bimetallic complex plotted against the LUMO energy of the metal fragment.}
    \label{fig:MOEs}
\end{figure*}

Lastly, after investigating the correlation between biradical character and TTF C-C bond length we aim to resolve the interaction between TTFtt$^{2-}$ ligand with the metal fragments. We reoptimized the geometries with spin-restricted DFT to obtain closed-shell geometries, reducing the variations in the structure of the TTFtt$^{2-}$ ligand arising from multi-reference effects, yielding R(C-C)s that vary by only 0.02 \AA \, across the surveyed systems. As expected the use of closed-shell singlet geometries yields lower biradical character and larger T-S gaps; however, the deviations from the broken-symmetry data are minor and the trends remain unchanged. The data are displayed in Table \ref{tab:fragments}. \\

\begin{table*}[h]
    \centering
    \begin{tabular}{cccccccc}
         & Fe & Pd & Pt & PtCF$_3$ & Ni & PtCOD & Sn  \\
         R(C-C) & 1.39 & 1.40 & 1.40 & 1.40 & 1.40 & 1.41 & 1.41 \\
         \hline
        $\Delta E_{T-S} \text{CASSCF}$ & 1.79 & 3.39 & 3.39 & 4.15 & 4.48 & 6.09 & 8.52 \\
        $\Delta E_{T-S} \text{NEVPT2}$ & 4.43 & 8.88 & 8.91 & & 11.98 & 16.55 & 20.29 \\
        $\lambda_{\text{LUNO, S}}$ & 0.74 & 0.62 & 0.61 & 0.56 & 0.55 & 0.45 & 0.33 \\
        $\lambda_{\text{HONO, S}}$ & 1.26 & 1.38 & 1.39 & 1.44 & 1.45 & 1.55 & 1.67  \\
        \hline
        $E_{HOMO, TTF}$ & 0.05923 & 0.05758 & 0.05782 & 0.05803 & 0.05937 & 0.05746 & 0.05488 \\
        $E_{LUMO,M}$ & -0.33538 & -0.38766 & -0.38249 & -0.40023 & -0.39087 & -0.51654 & -0.5215 \\
        $\Delta E_{L-H}$ & 10.738 & 12.116 & 11.981 & 12.470 & 12.252 & 14.056 & 15.684 \\
    \end{tabular}
    \caption{Data for the closed-shell singlet optimized complexes used in the fragment analysis. $\Delta E_{T-S}$ denotes the triplet-singlet gap \textcolor{black}{in kcal/mol} and the $\lambda$ denote the natural occupation numbers of the bimetallic complex. $E_{HOMO, TTF}$ and $E_{LUMO,M}$ denote the energy of the HOMO of the TTFtt$^{2-}$ fragment and the LUMO energy of the metal fragment, respectively, \textcolor{black}{in Hartree} while $\Delta E_{L-H}$ denotes the energy difference between these two orbitals \textcolor{black}{in eV}}.
    \label{tab:fragments}
\end{table*}

To investigate the role of the orbital interaction between the TTFtt$^{2-}$ ligand and the metal fragments in the determination of biradical character and T-S splitting, we perform DFT calculations on separate metal and TTFtt$^{2-}$ fragments in their closed-shell singlet geometry using the B3LYP functional and the larger def2-TZVP basis set for all atoms. The data are displayed in Table \ref{tab:fragments}. Analysis of the frontier MO energies of the individual fragments reveals the dominating interaction to be that between the metal fragment based LUMO and the TTFtt$^{2-}$ based HOMO. This interaction is illustrated in Figure \ref{fig:MOEs} (a). We define the number of unpaired electrons in the HONO and LUNO as $N_{ue} = 2-\lambda_{HONO} + \lambda_{LUNO}$, which gives a measure of the biradical character present in a given system, with $N_{ue}$ being equal to 2 in the case of a perfect biradical and 0 in the case of a closed-shell system. \\

Plotting the orbital energy gap between the metal fragment LUMO and the TTFtt$^{2-}$ HOMO obtained with DFT against $N_{ue}$ from [4,4] CASSCF calculations of the bimetallic complex yields a linear fit with $R^2 = -0.98$ (displayed in Figure \ref{fig:MOEs} (b)). Removing the variable of the TTFtt$^{2-}$ LUMO by plotting $N_{ue}$ against the metal fragment LUMO energy (Figure \ref{fig:MOEs} (c)) also reveals a linear dependence with a comparable $R$ value of 0.94. This suggests that, bar minor changes to the TTFtt$^{2-}$ orbitals arising from small structural distortions, the degree of biradical character in bimetallic TTFtt$^{2-}$ bridged complexes is determined by the LUMO energy of the terminal metal fragments. As the degree of fractional occupation correlates with the T-S gap, the same trends are observed for $\Delta E_{\text{T-S}}$ with a plot of $\Delta E_{\text{T-S, NEVPT2}}$ against the LUMO energy of the metal fragment shown in Figure \ref{fig:MOEs} (d), yielding $R^2 = -0.96$. This data excludes the PtCF3 complex whose size excluded it from NEVPT2 calculations. \\

While the computational data reveals a near-perfect linear dependence between $\Delta E_{L-H}$ of the fragments obtained with DFT and the observed degree of biradical character and $\Delta E_{\text{T-S}}$ calculated with high-level CASSCF and NEVPT2, inspection of the frontier orbitals shows that there is not a good symmetry match between the metal fragment based LUMO and the TTFtt$^{2-}$ HOMO. Inspection of the frontier metal fragment orbitals reveals the LUMO+2 to exhibit the correct $\pi$ symmetry to ideally interact with the TTFtt$^{2-}$ $\pi$ system; however unlike in the case of $\Delta E_{L-H}$, analysis of the orbital energies shows no clear correlation with the predicted NONs and T-S gaps. Another more plausible interaction for the trend in biradical character based on chemical intuition would be the interaction of the metal fragment based HOMO with the TTFtt$^{2-}$ LUMO, which would consequently lower the HOMO-LUMO gap in the bimetallic complex. However, inspection of this data again does not yield a good predictive model for the observed NONs and T-S gaps in the bimetallic complexes. \\

Our fragment analysis reveals a readily implemented one-parameter model that may be used to screen for new terminal ligands and metal centers for the development of bimetallic TTFtt$^{2-}$ bridged complexes with tunable biradical character. Metal fragments with high lying LUNOs yield strongly correlated complexes while those with low energy LUNOs yield closed-shell compounds. Consequently, earlier transition metals, such as Fe(II), yield complexes with strong biradical character and near-degenerate singlet and triplet states, while post-transition metals, such as Sn, yield closed-shell complexes with well separated singlet and triplet states. For a given metal center, the use of strong field ligands and those providing for greater electron delocalization and correspondingly greater splitting of the fragment frontier orbitals yields more fractional occupation of the frontier NOs in the bimetallic complexes, as illustrated by the series Pt $>$ PtCF3 $>$ PtCOD in order of decreasing biradical character. Finally, the biradical character introduced from orbital interactions between the metal and TTFtt$^{2-}$ fragments results in a synergistic  geometric change to the TTFtt ligand, with the central C-C bond length being shifted to shorter distances with increasing biradical character, as discussed in the previous section of this article. \\

\section{Conclusions}
We have performed high-level {\em ab initio} calculations on a series of recently synthesized and experimentally characterized bimetallic TTFtt-bridged complexes, demonstrating the presence of significant organic TTFtt-based biradical character in their electronic structure that varies with the identity of the metal fragments. To investigate the interplay between biradical character and structural changes to the TTFtt ligand, we performed CASSCF/ACSE calculations on CH$_2$ capped TTFtt$^{2+}$, revealing correlation between the degree of biradical character and the central TTF C-C bond length. Greater biradical character results in an increase in LUNO occupation accompanied by a corresponding decrease in HONO occupation; as the former is of $\pi$ bonding character across the central C-C atoms while the latter is $\pi$ antibonding, an increase in biradical character results in a shorter C-C distance. Comparison with experimental XRD data shows that measurement of this TTF core C-C distance provides a tool for the experimental determination of the multi-reference correlation and biradical character in a given compound and allows for the easy comparison of the biradical properties of TTFtt-bridged species. \\

Furthermore, we have performed orbital analyses on the different metal and TTFtt$^{2-}$ fragments comprising the various bimetallic systems, revealing a linear dependence of the T-S gap and biradical character of the bimetallic parent compounds on the energy spacing between the metal fragment LUMO and the TTFtt$^{2-}$ based HOMO. As the structural changes to the TTFtt$^{2-}$ ligand are minor compared to the MO changes in the metal fragments, this translates into a linear dependence on the energy of the metal fragment LUNO alone. Consequently, evaluation of fragment MO energies presents a simple one-parameter model for the prediction of biradical character and singlet-triplet splitting exhibited by bimetallic TTFtt bridged complexes, which may be evaluated with DFT at inexpensive computational cost. We demonstrate that increased biradical character may be achieved through the use of higher field ligands or earlier transition metals. Furthermore, since the character of this metal-fragment-based LUNO is largely M-L sigma star, this model provides a very general prediction that stronger sigma donor ligands should lead to more diradical character. \\

The results presented in this paper provide valuable insight for the intelligent design of future bimetallic TTFtt linked compounds with desirable properties. Using inexpensive DFT calculations to obtain the orbital energies of metal fragments provides a screening tool in the synthesis of complexes with strong TTFtt based organic biradical character which are prime candidates in the design of molecular qubits\cite{qubits, qubit} or switches\cite{FeSwitch, SwitchProcesses}, with possible applications in spintronics\cite{spintronics} or quantum information storage\cite{qistorage}. It may also form the basis for the development of future machine learning models for this purpose\cite{ML-TMComplex}. Finally, our calculations investigating the correlation between the central TTF C-C bond length and the observed biradical character provide valuable insight into the interplay between structural and electronic effects in this class of compounds and demonstrate that the TTF C-C bond length, which may be experimentally determined via XRD, presents a diagnostic quantity for the degree of organic biradical character present in a given TTFtt based system. \\

\textcolor{black}{The insight gained from this study informs the design of future TTFtt molecules. Specifically, these findings indicate synthetic chemists can manipulate the TTFtt electronic structure through the manipulation of capping metal orbital energies through selection of different capping ligands. For example, the donor properties of phosphorus-based ligands can be manipulated through different substituents\cite{Phos, Phos2}. Thus, these ligands are promising candidates for facile TTFtt electronic structure manipulation, as can be seen when comparing the diradical character and singlet-triplet gaps of Ptdppe and PtCF3.}\\

\begin{acknowledgement}
D.A.M. gratefully acknowledges support from the Department of Energy, Office of Basic Energy Sciences, Grant DE-SC0019215, the ACS Petroleum Research Fund Grant No. PRF No. 61644-ND6, and the U. S. National Science Foundation Grant No. CHE-1565638.
\end{acknowledgement}

\begin{suppinfo}
XYZ coordinates, as well as raw and computational data can be found in the SI.
\end{suppinfo}

\bibliography{citations}

\end{document}